
\input phyzzx
\hfuzz 50pt
\font\mybb=msbm10 at 12pt

\def\Bbb#1{\hbox{\mybb#1}}

\def\bR{\Bbb {R}}
\def\bE{\Bbb {E}}

\def\bfomega{\omega\kern-7.0pt \omega}


\REF\ati{M. Atiyah \& N. Hitchin, {\sl The Geometry and Dynamics
of Magnetic Monopoles}, Princeton University Press (1988), New Jersey.}
\REF\gipst { G.W. Gibbons, G.Papadopoulos
 \& K.S.Stelle,   {\sl HKT and OKT Geometries 
on Soliton Black Hole Moduli Spaces}, Nucl. 
Phys. {\bf B508} (1997) 623; hep-th/9706207.}
\REF\ast {A.Strominger,{\sl Open p-branes}, 
Phys. Lett. {\bf B383} (1996) 44; hep-th/9512059.}
\REF\gppkt {G.Papadopoulos \& P.K.Townsend,
{\sl Intersecting M-branes}, Phys. Lett. 
{\bf B380} (1996) 273; hep-th/9603087.}
\REF\jgp {J.Gutowski \& G.Papadopoulos,{\sl 
The moduli spaces of worldvolume brane solitons},Phys. Lett. {\bf
B432} (1998) 97; hep-th/9802186.}
\REF\blla {D.Bak,J.Lee \& H.Min,{\sl Dynamics of
 BPS  States in 
the Dirac-Born-Infeld Theory},hep-th/9806149.}
\REF\west{ N. Lambert \& P. West, {\sl Monopole Dynamics from
the M-fivebrane},  hep-th/9811025.}
\REF\witten{E. Witten, {\sl Bound States of
 Strings and p-Branes }, Nucl. Phys. 
{\bf B460} (1996) 335; hep-th/9510135.}
\REF\schw{J. Schwarz, {\sl An $SL(2,\bR)$ Multiplet of 
Type IIB Superstrings}, Phys. Lett. {\bf B360} (1995) 13: Erratum-ibid
{\bf B364} (1995) 152;
hep-th/9508143.}
\REF\gima {G.W.Gibbons \& N.S.Manton,{\sl 
The moduli space metric for well-separated BPS 
monopoles}, Phys. Lett. {\bf B356} (1995) 32; hep-th/9506052.}
\REF\cedw {M.Cederwall \& A.Westerberg,{\sl
 World-volume fields,SL(2,Z) and duality:the 
type IIB 3-brane},J. High Energy Phys. {\bf 02} (1998) 004; hep-th/9710007.}
\REF\kallosh{E. Bergshoeff, R. Kallosh, G. 
Papadopoulos \& T. Ortin, {\sl Kappa-symmetry
Supersymmetry and Intersecting Branes}, 
Nucl. Phys. {\bf B502} (1997) 149; hep-th/9705040.}
\REF\malda{C.J. Callan \& J. Maldacena, {\sl Brane Dynamics from the 
Born-Infeld
Action}, Nucl. Phys. {\bf B513} (1998) 198;  hep-th/9708147.}
\REF\gibbonse{G.W. Gibbons, {\sl Born-Infeld particles 
and Dirichlet p-branes}, Nucl. Phys. {\bf B514} (1998) 603; hep-th/9709027.}
\REF\fere {R.C.Ferrell \& D.M.Eardley,{\sl 
Slow motion scattering and coalescence of 
maximally charged black holes}, Phys. Rev.
Lett. {\bf 59} (1987) 1617.}
\REF\felsm {A.G.Felce \& T.M.Samols,{\sl 
Low-energy dynamics of string solitons}, Phys. 
Lett. {\bf B308} (1993) 30; hep-th/921118.}
\REF\shir {K.Shiraishi,{\sl Moduli space 
metric for maximally-charged dilaton black 
holes},Nucl. Phys. {\bf B402} (1993) 399.}
\REF\ggpt {J.P. Gauntlett, G.W.Gibbons, G.Papadopoulos
\& P.K.Townsend, {\sl Hyper-Kahler Manifolds
and Multiply Intersecting Branes}, Nucl. Phys.
 {\bf B500} (1997) 133.}
\REF\cpapad { R.A.Coles \& G.Papadopoulos,{\sl The 
geometry of one-dimensional supersymmetric 
nonlinear sigma models}, Class. Quantum Grav.
 {\bf 7} (1990) 427.}
\REF\ppd {G.Papadopoulos,{\sl T-duality and 
the worldvolume solitons of five-branes and
KK-monopoles},Phys. Lett. {\bf B434} (1998) 277; hep-th/9712162.}


\Pubnum{ \vbox{ \hbox{DAMTP-1998-159}\hbox{} } }
\pubtype{}
\date{November, 1998}
\titlepage
\title{The Dynamics of D-3-brane Dyons and Toric Hyper-K\"ahler Manifolds }
\author{J.Gutowski \& G.Papadopoulos}
\address{DAMTP, \break Silver Street, \break University of Cambridge,\break
 Cambridge CB3 9EW}



\def\C{\mkern1mu\raise2.2pt\hbox{$\scriptscriptstyle|$}\mkern-7mu{\rm C}}

\def\pd{\partial_}

\def\m{\mu}
\def\n{\nu}
\def\l{\lambda}
\def\a{\alpha}
\def\b{\beta}
\def\k{\kappa}
\def\p{\rho}
\def\s{\sigma}
\def\t{\tau}
\def\x{\chi}
\def\g{\gamma}
\def\f{\phi}
\def\d{\delta}
\def\e{\epsilon}

\def\tt{\theta}

\def\cL {{\cal {L}}}
\def\cF {{\cal {F}}}


\font\mybb=msbm10 at 12pt

\def\Bbb#1{\hbox{\mybb#1}}

\def\bR{\Bbb {R}}
\def\bE{\Bbb {E}}

\def\bZ {\Bbb{Z}}
\def\bfomega{\omega\kern-7.0pt \omega}


\abstract{We find the dyonic worldvolume solitons due to parallel (p,q) strings
ending on a D-3-brane.
These solutions preserve $1/4$ of bulk supersymmetry. Then we investigate the
scattering of well-separated dyons and find that their moduli space is a
toric hyper-K\"ahler manifold. In addition, we present the worldvolume solitons
of the D-3-brane which are related by duality to the M-theory configuration of
two orthogonal membranes ending on a M-5-brane. We show that these
solitons preserve $1/8$ of
 supersymmetry and compute their effective action.}


\chapter{Introduction}

The low energy scattering of solitons can be investigated 
using the moduli space approximation. One of the attractive
features of this picture is that moduli spaces are equipped
with novel geometries. 
For example, the moduli spaces of BPS 
monopoles are hyper-K\"ahler
manifolds [\ati] and the moduli spaces
of  (4+1)-dimensional black holes which preserve $1/4$ of supersymmetry
are hyper-K\"ahler manifolds with
 torsion [\gipst]. Other examples include the moduli spaces
 of some eight-dimensional
black holes which are octonionic
 manifolds with torsion.

In the last few years it has emerged
that the worldvolume solitons of
branes have a bulk interpretation. These solitons are due to brane
intersections or branes ending on other 
branes [\ast, \gppkt]. From the perspective of
one of the branes involved in such 
configurations, the intersection or
the brane boundary  is associated
 with the soliton 
of its worldvolume theory.  
The moduli spaces of worldvolume solitons 
of branes which preserve $1/4$
of bulk supersymmetry and involve a non-trivial
Born-Infeld field  have recently been 
investigated [\jgp, \blla, \west].
It has been found that some of their 
geometries are similar to those 
of the moduli spaces of supersymmetric black holes.

In this paper, we shall investigate
 the worldvolume dyons that arise from 
(p,q)-strings [\witten, \schw] ending on
a D-3-brane. The bulk configuration of a fundamental
 string along the direction $4$ and ending on a
D-3-brane, which lies in the directions $1,2,3$,  is
$$
\eqalign{
D3&:\, 0,1,2,3,*,*,*,*,*,*
\cr
F1&:\, 0,*,*,*,4,*,*,*,*,*\ .}
\eqn\mone
$$
Acting on the above configuration with the  
 U-duality group  $SL(2, \bZ)$ of type IIB strings,
the D-3-brane remains invariant but the fundamental 
string turns into a (p,q) string.
It is known that the above bulk configuration 
preserves $1/4$ of supersymmetry.
From the perspective of the D-3-brane, the string boundary 
is associated with a dyonic worldvolume
soliton that carries (p,q) charge and
 preserves $1/4$ of supersymmetry.
A related bulk configuration is that of 
two orthogonal strings, one 
of which is fundamental
and the other is a D-string, ending possibly 
at different points on a D-3-brane. 
The bulk configuration is
$$
\eqalign{
D3&:\, 0,1,2,3,*,*,*,*,*,*
\cr
F1&:\, 0,*,*,*,4,*,*,*,*,*
\cr
D1&:\, 0,*,*,*,*,5,*,*,*,*\ .}
\eqn\fdthree
$$
This can be derived from the M-theory configuration of 
a M-5-brane and two orthogonal membranes
intersecting on a 0-brane by first reducing to ten
 dimensions and then using 
IIA/IIB T-duality. This supergravity solution
preserves $1/8$ of supersymmetry. If the  
fundamental string and the
D-string end at the same point in D-3-brane, the 
associated worldvolume soliton
is a $(1,1)$-dyon, i.e. a dyon of unit electric 
and  unit magnetic charges,
 preserving $1/4$
of bulk supersymmetry. This worldvolume soliton is also
associated with a (1,1)-string ending on the D-3-brane. More 
generally, if $q$ fundamental
strings and
$p$ D-strings all end at the same point in D-3-brane, the 
associated worldvolume solitons
are $N$ $(p',q')$-dyons, where   $N$ is the maximal common
divisor of $p,q$. This dyon is also
associated with $N$ $(p',q')$-strings ending 
on the D-3-brane. The supersymmetry 
preserved is $1/4$.
More general IIB configurations
can be found by acting on \fdthree\ with type IIB U-duality.

The effective theory of $n$ coincident D-3-branes 
in the linearized limit is N=4
super-Yang Mills with gauge group $U(n)$. This
theory is conjectured to have an $SL(2,\bZ)$ duality. This is
required in order to describe 
the low energy dynamics of the D-3-brane.
The worldvolume solitons associated
with a (p,q)-string ending on the 
D-3-brane are BPS (p,q)-dyons. The moduli
spaces of these dyons are hyper-K\"ahler 
manifolds. For well-separated dyons,
the moduli space becomes a toric hyper-K\"ahler manifold [\gima].
In this approximation, the forces due to massive intermediate
gauge bosons are suppressed. 
More recently a manifestly
 (non-linear) Dirac-Born-Infeld-type of 
action has been proposed for a
D-3-brane in [\cedw]. This action is 
kappa-symmetric and its field
equations are manifestly $SL(2, \bR)$ invariant. 
This invariance is achieved
 by using two
$U(1)$ gauge fields which  transform as a doublet 
under $SL(2, \bR)$ and then imposing a
self-duality condition to reduce the physical 
degrees of freedom due to gauge
fields to two. This non-linear action can also be used to
investigate the (p,q)-dyons on the D-3-brane. Since in this
action the intermediate massive gauge bosons are suppressed,
the moduli spaces of these dyons are expected to be toric
hyper-K\"ahler manifolds.

In this paper, we shall use the non-linear 
$SL(2, \bZ)$-invariant D-3-brane action
of [\cedw] modified appropriately by adding 
source terms to find the 
worldvolume solutions that are associated
with both the above bulk configurations.
 In particular, we shall find
dyonic configurations that have the
interpretation of $N$ parallel (p,q)-strings 
ending on the D-3-brane. We shall also find
solutions that have the interpretation of
 orthogonal fundamental strings and D-strings 
ending at different points on the
D-3-brane.
 We shall show, using the  kappa-symmetry 
transformation of the D-3-brane and its relation
to  supersymmetric configurations found in [\kallosh],
that these dyons preserve $1/4$ and 
$1/8$ of the bulk supersymmetry,
 respectively. 
Then we shall use the $SL(2, \bR)$-invariant action 
to show that the moduli 
spaces of $N$ worldvolume well-separated
dyons that preserve $1/4$ of supersymmetry 
are toric hyper-K\"ahler
manifolds. In addition, we shall give 
the effective action of some of
the worldvolume solutions that 
preserve $1/8$ of spacetime supersymmetry.

The paper is organized as follows. In section two, 
we present the appropriate static solutions of the 
$SL(2,{\bZ})$-invariant
action. In section three, we show that our classical solutions
preserve either $1 \over 4$ or $1 \over 8$ of
bulk supersymmetry using 
$\k$-symmetry. In section four, we review and improve certain aspects
of the computation of the metric on moduli space of  
0-brane worldvolume
solitons of D-p-branes that we shall use here. In section five, we find 
that the moduli space
of solutions preserving $1/4$ of
 supersymmetry is a toric
hyper-K\"ahler manifold.
In section six, we find the metric on the
 moduli space of some of
the configurations 
that preserve $1/8$ of supersymmetry. Finally, in section 
seven we give our
conclusions.

\chapter{The $SL(2,{\bZ})$ invariant action and static solutions.}

To find the   worldvolume dyons of D-3-brane, we shall use
the $SL(2,\bZ)$ covariant 
 action  of [\cedw] and choose 
as a IIB supergravity
background the ten-dimensional Minkowski spacetime.
  We also introduce
the two-form field strength
$$
 \cF_{\m\n}\equiv U^r (F_r)_{\m\n}=
U^r\big(\pd{\m} A_{ r\n} -\pd{\n}A_{ r\m} \big)
\eqn\mtwo
$$
where $U^r$ are complex constants satisfying 
$$
{i\over 2} \epsilon_{rs} U^r \bar U^s=1\ ,
\eqn\mthree
$$
and $A_r$ are one-form gauge
 potentials; $r=1,2$ and $\m,\n=0,...3$.
Under $A\in SL(2,\bR)$, the constants 
$$
U^T=(U^1, U^2)
\eqn\mfour
$$
 and field strengths
$$
F=\pmatrix{F_1\cr F_2}
\eqn\mfive
$$ 
transform as  $U^T\rightarrow U^T A^{-1}$ 
and $F\rightarrow A F$. So the two-form field
strength $\cF=U^T F$ is
$SL(2,\bR)$ invariant.
The constant IIB supergravity axion 
$\rho$ and dilaton $\phi$ are given
in terms of $\{U^r\}$ as
$$
\tau\equiv \rho+i e^{-\phi}={U^2\over U^1}\ .
\eqn\msix
$$
Under $SL(2,\bR)$, $\tau $ transforms as usual with
 fractional linear transformations. 
The bosonic part of the D-3-brane action [\cedw] is
$$
S_{D3}=\int d^{4}x\, \l\, 
\sqrt{-{\rm det}(g)}(1+{1\over 2}\cF 
\cdot\bar{\cF} -{1 \over 16}(\cF \cdot {}^*\cF)(\bar{\cF}
 \cdot {}^*\bar{\cF})+{1 \over 4}F\cdot F)
\eqn\pola
$$
where here $\l$ is a Lagrange multiplier,
 $F=dW$ is a 4-form field strength and $g$ is 
the induced worldvolume metric\foot{We  are 
using the conventions for the inner product 
$\phi\cdot \psi={1 \over p!}
\phi^{\m_1 ... \m_p}\psi_{\m_1 ... \m_p}$ and the dual form
 ${}^*\phi^{\m_0 ... \m_{3-p}}=$
${1 \over p! \sqrt{-g}}\e^{\m_0
... \m_{3-p} \m_{3-p+1} ... \m_{3}}
\phi_{\m_{3-p+1} ... \m_{3}}$, where $\e^{0123}=1$.}. This action is 
supplemented with a \lq self-duality'-like  relation  
$$
{i \over 2}({}^*F){}^*\cF=\cF-{1 \over 4}
(\cF \cdot {}^*\cF){}^*\bar{\cF}\ .
\eqn\sdr
$$
The duality relation has the status 
of a field equation
and should not be substituted into the
 action. We remark that in the static
 gauge the action \pola\
has  six scalars which describe the position 
of the D-3-brane in the ten-dimensional
Minkowski spacetime. Both 
that action and the self-duality
condition are invariant under 
the $U(1)$ transformation
$$
\cF\rightarrow e^{i\theta} \cF\ ,
\eqn\uone
$$
where $\theta$ is a angle.

There are four in total  electric and  magnetic
 charges associated with 
$F_1$ and $F_2$. However due to the self-duality condition, only two are 
independent. We shall choose
$$
\eqalign{
q&={1\over 4\pi}\int_{S^2} F_1
\cr
p&={1\over 4 \pi}\int_{S^2} F_2\ .}
\eqn\mseven
$$

We shall investigate dyon configurations for 
which the non-vanishing
fields are $F, \cF, \l$ and two
transverse scalars $X, Y$. In addition, 
since $\cF$ is $SL(2,\bR)$ invariant,
we can set without loss of generality
$$
\eqalign{
U^1&=1
\cr
U^2&=i\ ,}
\eqn\choic
$$
after possibly using a $SL(2,\bR)$ transformation 
and so $\cF=F_1+iF_2$. This choice
corresponds to the
$\tau=i$ vacuum of supergravity theory.  A
$U(1)$ subgroup of $SL(2, \bR)$ preserves \choic.

A solution to the field equations of \pola\ is  
$$
\eqalign{
\cF_{on}&=\pd{n}Z
\cr 
\cF_{pq}&=i \e_{pqr}\d^{rl}\pd{l}Z
\cr
X&=H_1
\cr
Y&=H_2\ ,}
\eqn\sol
$$
where $H_1, H_2$ are harmonic functions in $\bR^3$, i.e.
$$
\eqalign{
H_1&=m_1+\sum^N_{i=1} {\alpha_i\over |{\bf{x}}-{\bf{y}}_i|}
\cr
H_2&=m_2+\sum^{N'}_{i'=1} 
{\beta_{i'}\over |{\bf{x}}-{\bf{y}}'_{i'}|}
\cr
Z&=H_1-iH_2\ ,}
\eqn\meight
$$
and $m,n,p,q, r, l=1,2,3$. The rest of the fields are 
easily determined. For example,
$$
{}^*F = (2+\delta^{mn} 
\partial_mZ \partial_n\bar Z )\ ,
\eqn\brk
$$
where the Hodge star operation is taken with respect to the flat metric
on $\bE^(3,1)$.
The macroscopic bulk interpretation of this
solution is as $N$ fundamental- and 
$N'$ D-strings ending at
the points $\{{\bf{y}}_i; i=1,\dots, N\}$ and 
$\{{\bf{y}}'_{i'};i'=1,\dots, N'\}$ 
of the D-3-brane, respectively.
The charges of this solution 
are $p_i=\a_i$ and $q_{i'}=\b_{i'}$

Next let us suppose that $N=N'$ and 
${\bf{y}}_i={\bf{y}}'_{i}$.
Such solution describes $N$ worldvolume dyons and
 has the macroscopic bulk interpretation of $N$
dyonic strings with charges
 $\{(p_i,q_i)=(\a_i, \b_i); i=1,\dots, N\}$
ending at the points
 $\{y_i; i=1,\dots,N\}$ of the D-3-brane.
These strings are not parallel in 
the $X,Y$ plane but end orthogonally
on the D-3-brane. The angle
that these strings lie relative to  $X,Y$ directions is
$$
\tan\phi_i={\b_i\over \a_i}\ .
\eqn\mnine
$$  
We shall show in the next section that when all the 
strings are parallel the solution
preserves $1/4$ of supersymmetry. 
The mass of the dyons that we have found is infinite
since the length of the associated (p,q)-strings is infinite
(for a similar discussion see [\malda]). Therefore to compute
the metric on the moduli space of such solutions, we have to 
\lq regularize' the above solutions. This will be explained
in detail in section four.

To explore the $SL(2,\bR)$ 
invariance of the field equations
of our D-3-brane solution and demonstrate 
further their dyonic nature, let us take $H_1=0$
and
$$
H=H_2=1+{\beta\over |\bf{x}|}\ .
\eqn\mten
$$ 
For such solution, the electric part of $F_1$ and 
the magnetic part of $F_2$ vanish. So it
describes a particle
 of charge $(0, \beta)$ in the
$\tau=i$ vacuum. We can now use the 
method in [\schw], to construct
a dyonic solution with charge $(p,q)$ at
 any vacuum $\tau$. For this we take
$H$ as above for some $\beta$  parameter 
and set $f=F_1$ for the non-vanishing
two-form field strength. Then
we act on the magnetic part of  $(F_1, F_2)=(f, 0)$ with 
the transformation
$$
A={1\over \sqrt{p^2+q^2}}\pmatrix{q& -p\cr p&~   q}
\eqn\meleven
$$
of $U(1)\subset SL(2, \bR)$ that stabilizes 
the vacuum $\tau=i$ to give
$$
\eqalign{
f_1&={q\over \sqrt{p^2+q^2}} f
\cr
f_2&={p\over \sqrt{p^2+q^2}}f\ .}
\eqn\none
$$
Now if we choose $\beta=\sqrt{p^2+q^2}$, then the solution with magnetic
part
 $(F_1,F_2)=(f_1,f_2)$ describes
a dyon with charges (p,q) in the $\tau=i$ vacuum. The electric part of
$(F_1,F_2)$ can be easily determined in a similar way.
To give the solution of a (p,q)-dyon at a generic vacuum
$\lambda_0$, we first write for the magnetic part of  
$(F_1,F_2)=(f_1, f_2)= (\cos\theta f, \sin\theta f)$,
where
$$
e^{i\theta}={q+ip\over \sqrt{p^2+q^2}}\ .
\eqn\ntwo
$$
Then we act with the
transformation
$$
A=\pmatrix{ e^{-{\phi_0\over2}}& -\rho_0 e^{{\phi_0\over2}}\cr
0&~e^{{\phi_0\over2}}}\ .
\eqn\nthree
$$
The vacuum $\tau=i$ is transformed to  
$\tau_0=\rho_0+ie^{-{\phi_0}}$. 
Moreover the magnetic part $({\bf B}_1, {\bf B}_2)$  of  $(F_1, F_2)$ is 
$$
\eqalign{
{\bf B}_1&=e^{-{\phi_0\over2}} 
\cos\theta f-\rho_0 e^{{\phi_0\over2}} \sin\theta f
\cr
{\bf B}_2&= e^{{\phi_0\over2}} \sin\theta f\ .}
\eqn\nfour
$$
For this dyon to have charge $(p,q)$, we set
$$
e^{i\theta}= {(p\lambda_0+q) 
e^{{\phi_0\over2}}\over \sqrt{(p \rho_0+q)^2 e^{\phi_0}
+p^2 e^{-{\phi_0}}}}
\eqn\nfive
$$
and 
$$
\beta=\sqrt{(p \rho_0+q)^2 e^{\phi_0}
+p^2 e^{-{\phi_0}}}\ .
\eqn\nsix
$$
The electric parts of $(F_1, F_2)$ can be easily
 determined in a similar way.

After quantization, the charges $(p,q)$ and $(p',q')$ of two
worldvolume dyons are 
quantized according to the 
Schwinger-Zwanziger quantization condition
$$
pq'-qp'\in \bZ\ .
\eqn\nseven
$$
Then the bulk interpretation of our dyonic solution
is that of $\{ k_i; i=1,\dots, N\}$ 
$\{(\tilde\alpha_i, \tilde\beta_i)\}$-strings
ending on the D-3-brane, 
where $k_i$ is the maximal 
common divisor of $(\alpha_i, \beta_i)$.

\chapter{$\k$-symmetry of D-3-brane dyons}

To find the supersymmetry preserved 
by a worldvolume solution of 
a brane, we make use of the kappa-symmetry transformations
$$
\delta\tt=P_+\kappa\equiv (1+\Gamma)\kappa\ ,
\eqn\neight
$$
where
$\kappa$ is the kappa-symmetry parameter, $\Gamma$ 
satisfies $\Gamma^2=1$ and
${\rm Tr}\Gamma=0$. Then  $P_\pm=(1/2)(1\pm\Gamma)$ are 
projection operators. Then it has been shown in [\kallosh] that the
supersymmetry condition is
$$
(1-\Gamma)\epsilon=0\ ,
\eqn\kone
$$
where $\epsilon$ is the spacetime supersymmetry parameter.

To describe the kappa-symmetry 
transformations of D-3-brane, we introduce
$$
\gamma_{(k)}={1\over k!} dX^{A_1}\wedge \dots\wedge  
dX^{A_k} \Gamma_{A_1}\dots \Gamma_{A_k}\ ,
\eqn\lone
$$
where $X$ are the embedding maps of the D-3-brane
 worldvolume into spacetime and $\{\Gamma_{A}; A=0,\dots
9\}$ are the spacetime gamma-matrices in the 
Majorana-Weyl representation. Then for the D-3-brane, the
projections
$P_\pm$ are
$$
({}^*F) P_{\pm} \eta={1 \over 2}({}^*F)\eta \mp {i \over 2}{}^*
(\cF \wedge \g_{(2)}) {\bar{\eta}} \pm i {}^*(\g_{(4)}) \eta\ ,
\eqn\arl
$$  
where $\eta=\eta_1+i\eta_2$; $\eta_1, \eta_2$ are 
16-component Majorana-Weyl
spinors of the same chirality.
The supersymmetry condition is
$$
{1 \over 2}({}^*F)\epsilon - {i \over 2}{}^*
(\cF \wedge \g_{(2)}) 
{\bar{\epsilon}} - i {}^*(\g_{(4)}) \epsilon=0\ .
\eqn\hone
$$
For our solutions, this condition can be written as
$$
*(Im(\cF) \wedge \g_{(2)}) \e_1 - (2*(\g_{(4)})+
*(Re(\cF) \wedge \g_{(2)})) \e_2 =
 (1+{1 \over 2}|\nabla X|^2 +{1 \over 2}|\nabla Y|^2) \e_1 \ ,
\eqn\arq
$$
where $\epsilon=\epsilon_1+i \epsilon_2$.
If the exterior derivatives $dX, dY$ of the fields $X,Y$ of 
our solution are linearly independent, then \arq\
implies that
$$
\eqalign{
\e_1 &= -\Gamma_0 \Gamma_1 \Gamma_2 \Gamma_3 \e_2
\cr
\e_1 &= \Gamma_1 \Gamma_2 \Gamma_3 \Gamma_5 \e_1
\cr
\e_1 &= \Gamma_0 \Gamma_4 \e_1\ ,}
\eqn\art
$$
and so the solution preserves $1/8$ of bulk supersymmetry.
If on the other hand 
$$
dX-r dY=0\ ,
\eqn\lirel
$$
where $r$ is a real number,
 then \arq\ implies that
$$
\eqalign{
\e_1 &= -\Gamma_0 \Gamma_1 \Gamma_2 \Gamma_3 \e_2
\cr
-\e_1 &= \big(r (\Gamma_0+ \Gamma_4)+ \Gamma_5\big) 
\Gamma_1 \Gamma_2 \Gamma_3  \e_1\ ,}
\eqn\art
$$
and the solution preserves $1/4$ of bulk 
supersymmetry. It is clear
that for such solutions $N=N'$, 
${\bf{y}}_i={\bf{y}}'_i$ and $\alpha_i \beta_j=
\alpha_j \beta_i$ for every $(i, j)$. The supersymmetry  
preserved by our dyons is
in agreement with that preserved by the bulk 
configurations in the introduction for
which all strings are parallel. 
We can identify $r$ with  $\cot\phi$, where $\phi$ is the angle
that the strings lie in the $X,Y$ plane. We can 
always take $r=0$ or $r=\infty$ 
using possibly a field
redefinition, i.e. a rotation in the $X,Y$ plane.

\chapter{The Moduli Spaces of 0-brane Worldvolume Solitons Revisited}

Before we compute the metric on 
the moduli space of the dyons
we have found in the previous 
sections, we shall briefly review
and improve on some aspects
of the computation we have done 
for the metric on the moduli
space of 0-brane  worldvolume solitons of D-p-branes in [\jgp]. 
The non-vanishing fields for such solution is a transverse scalar $Y$
and the electric part of the Born-Infeld 
field $F$ of the D-4-brane. The solution [\malda, \gibbonse] is
$$
\eqalign{
F&= dt\wedge dH
\cr
Y&=H\ ,}
\eqn\zerobrane
$$
where $H$ is a harmonic function in $\bR^4$. We may take
$$
H=1+\sum_{i=1}^N {\mu_i\over |x-y_i|^{p-2}}\ ,
$$
where $\{x^a; a=1,\dots, p\}$ are the spatial worldvolume
coordinates of a D-4-brane and 
$\{(y^a_i; a=1,\dots,p); i=1,\dots, N\}$
are the locations on $N$ 0-brane 
worldvolume solitons. The location of the
D-p-brane at $|x|\rightarrow \infty$ is at $Y=1$.

Next let us consider the case of one such
 0-brane, i.e. $N=1$. We can
choose without loss of generality $y_1=0$ and 
so the 0-brane is located
at $|x|=0$.
 It has been demonstrated in [\malda] that this solution
 has  infinite mass. For
this, the energy $E$ of the configuration 
is computed in a region $|x|>R$
and it is found that $E(R)\sim |Y(R)-1|$. This is
precisely the mass $M(R)$ of a string with constant 
tension and length $|Y(R)-1|$.  Now
 it can be easily seen from \zerobrane\ that as
$R\rightarrow 0$, $Y(R)\rightarrow \infty$ and so
$E\rightarrow \infty$. Therefore this infinity has
 the physical interpretation
as the mass of a  string with constant tension and infinite length.
The introduction
 of the characteristic
size $R$ for the 0-brane can be thought as a cut off for 
distances close to its location.

For the  computation of the moduli space metric, we should keep
the masses of the 0-branes finite. This, as we have seen, requires
a  \lq regularization' of the  solution \zerobrane. However, the cut 
off regularization described above
 is not convenient. A more instructive way to \lq regularize' the solution
 is to
assume that the 0-brane solitons are balls of radius $R$ with
centres at
 $\{y_i; i=1,\dots,N\}$ and
with constant charge density $\{\rho_i; i=1,\dots, N\}$, where
$$
\rho_i=\cases{0\qquad ~~\,: |x-y_i|>R\cr
{p\mu_i\over R^p}\qquad : |x-y_i|<R}\ .
\eqn\zone
$$
The regularized 0-brane solution is as in  \zerobrane\ but now
$H=H(R)=1+\sum_{i=1}^NH_i$, where
$$
H_i=\cases{{\mu_i\over |x-y_i|^2}\qquad \quad 
\qquad \qquad \qquad ~: |x-y_i|>R
\cr
	{p \mu_i\over 2 R^{p-2}}+{(2-p)\mu_i\over 2 R^p} 
|x-y_i|^2 \qquad : |x-y_i|<R}\ .
\eqn\ztwo 
$$
One of the advantages of this regularization is that 
the fields are continuous
at $|x-y_i|=R$ and so certain surface terms due to partial 
integrations that appear in
the calculation of the moduli metric can be easily handled.
The energy of a single 0-brane is 
$$
E={2p\Omega_{p-1}\mu\over p+2}  (Y(R)-1)
\eqn\zthree
$$
and so again it has the interpretation of the mass of a
string with constant tension $T_f={2p\Omega_{p-1}\mu\over p+2}$ 
of length $Y(R)-1$, where $\Omega_{p-1}$ is 
the volume of a $(p-1)$-sphere
of unit radius and $\mu=\mu_1$.

The moduli calculation can proceed as in [\jgp]  by  using the
linearized\foot{This calculation can also be done with the full
non-linear Born-Infeld Lagrangian.} 
p-brane action
$$
S= S_0+S_{Source}+S_{free}\ ,
\eqn\zfour
$$
where
$$
S_0={1\over2} \int d^{p+1} x 
\big(\eta^{\mu\nu} \partial_\mu Y\partial_\nu Y+{1\over2}
F_{\mu\nu} F^{\mu\nu}\big)\ ,
\eqn\zfive
$$
$$
S_{Source}=(2-p) \Omega_{p-1}\sum_{i=1}^N 
\int d\tau_i \rho_i \big( Y+ A_\mu {\partial
y_i^\mu\over \partial \tau_i}\big)
\eqn\zsix
$$
and
$$
S_{free}=-\sum_{i=1}^N M_i \int d\tau_i+
(p-2) \Omega_{p-1} \sum_{i=1}^N \mu_i \int d\tau_i\ ;
\eqn\zseven
$$
$ \{M_i; i=1,\dots,N\}$ are the masses of the worldvolume 0-branes.
 The sources $S_{Source}$ that we have added in the action
 are those of the classical solution
\zerobrane. These sources should have a bulk interpretation 
as a fundamental string and
a D-p-brane; the former is the source of the 
Born-Infeld field $F$ while the latter is
the source of the scalar $Y$. The first part in $S_{free}$
is the action of $N$ free relativistic particles with masses
$\{M_i; i=1,\dots,N\}$. The second part in $S_{free}$ has been added
in order to make the effective action of the worldvolume 0-branes
independent from the (asymptotic) position of the D-p-brane.

Repeating the computation as in [\jgp] but now with the
\lq regularized' solution, we find that the metric on the
moduli space is
$$
ds^2=\sum_{i=1}^N M_i |dy_i|^2+ (p-2) \Omega_{p-1} 
\sum_{i<j}^N \mu_i \mu_j  k(y_i-y_j) |dy_i-dy_j|^2\ ,
\eqn\modmetricp
$$
where
$$
k(y_i-y_j)={p\over R^p}\int_{S^{p-1}}\int_0^R {r^{p-1}\over
|r+y_i-y_j|^{p-2}} dr d\Omega_{p-1}\ .
\eqn\znine
$$
All the properties of the moduli metric 
found in [\jgp] are still valid
provided that $|y_i-y_j|>R$. For example, 
the moduli space of
0-brane solitons on the D-4-brane is a 
hyper-K\"ahler manifold with torsion.
The same applied for the moduli space of the 
worldvolume self-dual string soliton of
M-5-brane. Now if the separation distance of 
worldvolume 0-branes is much larger than
the size of the balls, $|y_i-y_j|>>R$, then
$$
k(y_i-y_j)= {1\over |y_i-y_j|^{p-2}}+ O({R\over|y_i-y_j| })\ .
\eqn\zten
$$
Substituting this into \modmetricp, we 
recover the result found in [\jgp]. 
It is clear then that the quadratic in the 
velocities interaction terms of well-separated  0-brane solitons
are independent of the regularization 
of their masses. Therefore, the moduli metric found above 
is the moduli metric of well-separated 
0-brane solitons. The order parameter is the ratio of
the size of the 0-branes with their 
separation distance. This approximation 
is the same as that employed
in [\gima] to find that the asymptotic metric of the BPS monopoles
moduli space is toric hyper-K\"ahler. This way 
of understanding the moduli
metric \modmetricp\ will be further enforced 
by the result of the next
section. Incidentally, the large separation 
distance approximation
is consistent with suppressing the interactions 
of the 0-branes due the massive
intermediate gauge bosons of the Coloumb 
branch of the full non-abelian
effective theory of  D-p-branes. This is because 
the forces of such interactions
are short range and so in the above limit do 
not contribute. However
for the full treatment of the moduli space of 
D-0-branes, they should be taken
into account.  

\chapter{Moduli Space metrics}

To determine the metric on the moduli 
space of our dyons, we 
have to compute the
quadratic term in the  velocities of their effective
action. For this we should again \lq regularize'
our solution in section two using one of the methods that
we have described  in the previous section for the other 
0-brane worldvolume solitons. To
simplify matters, we shall perform our computations 
assuming that the dyons are well-separated.
We shall
investigate the moduli
 space for those solutions
for which $N=N'$ and ${\bf{y}}_i={\bf{y}}'_i$.
 For this using the method of 
[\fere,\felsm, \shir, \jgp],  we add  to
the action \pola\ the source term
$$
\eqalign{
S_{\rm {source}}=& -{4 \pi \s}\sum_{i=1}^{N} \int d \t_i   
\d({\bf{x}}-{\bf{y}}_i)
({\hat{p}}_i X + {\hat{q}}_i Y + {1 \over 2}p_i 
(A_1)_{ \m}{d {y_i}^{\m} \over d \t_i}-{1 \over 2}q_i 
(A_2)_{ \m}{d {y_i}^{\m} \over d \t_i})
\cr &
 + {\s \over 2}
\int d^{4}x ({1 \over 96}\e^{\m \n \p \s} F_{\m \n \p \s}-2) \ ,}
\eqn\arg
$$
 where $\sigma$ is a constant and 
$\{\tau_i; i=1,\dots, N\}$ 
are the  proper times of dyons.  The scalar 
charges $\{{\hat{p}}_i, {\hat{q}}_i\}$
can be chosen to be independent from $\{p_i, q_i\}$. 
Though they can be related
for particular solutions. For the purpose 
of this paper, we take
$$
\eqalign{
{\hat{p}}_i &= \sqrt{{p_i}^2 + {q_i}^2} \cos \f_i
\cr
{\hat{q}}_i &= \sqrt{{p_i}^2 + {q_i}^2} \sin \f_i\ ,}
\eqn\hatpq
$$
where $\{\f_i; i=1,\dots, N\}$ are some phase angles. 
Note that the last two terms in \arg\  do not
affect the  field equations.

The parameters of the source terms and 
those of the solution \sol\ 
are related as
$$
\eqalign{
p_i& = {\hat{p}}_i=\a_i
\cr
q_i& = {\hat{q}}_i=\b_i\ .}
\eqn\ppqq
$$
These are the charges of the dyons. The expression \ppqq\ for the
scalar charges is consistent with that of 
\hatpq, if we choose the phase
angles for this solution as
$$
\tan\phi_i={\b_i\over \a_i}\ .
\eqn\uone
$$

The field equations of \pola\ with 
source terms \arg\ for the fields $\l, W, A$ 
are
$$
\eqalign{
{}^*F&=2\sqrt{1+{1\over 2}\cF \cdot\bar{\cF} -{1 \over 16}
(\cF \cdot {}^*\cF)(\bar{\cF} \cdot {}^*\bar{\cF})}
\cr
\l\, {}^*F&=\s
\cr
 \pd{\m}( \sqrt{-g}{}^*{\cF}^{\m \n})&=
4 \pi i \sum_{i=1}^{N}(p_i-iq_i) \d ({\bf{x}}-{\bf{y}}_i) 
{d {y_i}^{\n} \over d \t_i}\ ,}
\eqn\brc
$$
respectively. The field equations for the two 
transverse scalars $X,Y$ are
$$
\eqalign{
\pd{\n}(\l \sqrt{-g}({\cF^{(\m}}_{\a}\bar{\cF}^{|\a| \n)}
 +{1 \over 2}(\cF.\bar{\cF})&g^{\m \n}-
\cr
{1 \over 12}
{F^{\m}}_{\s \l \t}F^{\n \s \l \t})\pd{\m}X)&=
-4 \pi \s \sum_{i=1}^{N}{\hat{p}}_i \d ({\bf{x}}-{\bf{y}}_i)
\cr
\pd{\n}(\l \sqrt{-g}({\cF^{(\m}}_{\a}\bar{\cF}^{|\a| \n)}
 +{1 \over 2}(\cF.\bar{\cF}) & g^{\m \n}-
\cr
{1 \over 12}
{F^{\m}}_{\s \l \t}F^{\n \s \l \t})\pd{\m}Y)&=
 -4 \pi \s \sum_{i=1}^{N}{\hat{q}}_i \d ({\bf{x}}-{\bf{y}}_i).} 
\eqn\brh
$$
For the Hodge star operation and for the raising
 and lowering of indices we have used
the induced metric on the D-3-brane.  The
constant $\s$ is related to the tension of the D-3-brane.

We shall consider perturbations of  our solution that
involve the fields and the various charges. 
To illustrate how such
a perturbation can be done, let 
${\cal L}(\phi, s)$ be a Lagrangian
of a field $\phi$ and depending on 
some parameters $s$. Suppose that
we have a solution $(\phi_0, s_0)$ of 
this system. Next consider
a perturbation $(\phi(u), s(u))$ of this theory such that 
$(\phi(0), s(0))=(\phi_0, s_0)$. It is easy to see that
if
$$
{\partial {\cal L}(\phi, s)\over \partial s}|_{s=s_0}=0\ ,
\eqn\sfe
$$
then the linear terms in the perturbation
 of the Lagrangian vanish subject
to the field equations\foot{In the supergravity
 context, \sfe\ should be thought
as part of the supergravity field 
equations for the IIB dilaton and axion.} of
$\phi$. In addition,  it turns out that
only the linear terms in the perturbation 
of the fields and of the parameter
$s$ contribute
in the quadratic perturbation of the Lagrangian.

Next, we consider the following perturbation: 

\item{(i)} We allow 
$$
{\bf y}_i\rightarrow {\bf y}_i(t) \ .
\eqn\utwo
$$

\item{(ii)}The charges are perturbed as

$$
\eqalign{
\a_i \rightarrow p_i&=\a_i + \b_i \x_i
\cr
\b_i \rightarrow q_i&=\b_i - \a_i \x_i\ ,}
\eqn\cri
$$
where $\chi_i$ are the perturbations of the charges.

\item{(iii)} The perturbation of the transverse scalars $X,Y$
is induced only by that in ${\bf y}_i$, i.e
$$
\eqalign{
X(\alpha_i, {\bf y}_i)&\rightarrow X(\alpha_i, {\bf y}_i(t)) 
\cr
Y(\beta_i, {\bf y}_i)&\rightarrow Y(\beta_i, {\bf y}_i(t))}\
\eqn\scalarp 
$$ 
and so $X,Y$ are not perturbed linear in the 
velocities.

\item{(iv)} The perturbation in the vector 
potentials $A_1,A_2$ of $(F_1,F_2)$ is as
$$
\eqalign{
A_1(\alpha_i, \beta_i, y_i)& 
\rightarrow A_1(p_i, q_i, y_i(t))+ B_1
\cr
A_2(\alpha_i, \beta_i, y_i)& 
\rightarrow A_2(p_i, q_i, y_i(t))+ B_2\ ,}
\eqn\gaugep
$$
where $B_1, B_2$ are determined by the field equations.

The perturbations  of the scalar 
charges $\hat p_i, \hat q_i$ are 
determined by \hatpq. But in order 
to perturb the scalar fields as in
\scalarp\ in a way  consistent with the field equations,  the
phase angles $\phi_i$ remain unperturbed. 
Their values are those of the original
solution. The perturbations of $F$ and 
$\lambda$ are determined by those of the
other fields above using the field equations. 

To find $B_1, B_2$, we  consider the  fields equations of ${\cal F}$ 
both linearized in the fields
and the velocities. These are
$$
\eqalign{
\pd{0}(A_1)_n-\pd{n}(B_1)_{0}&={\e_n}^{ls}\pd{l}(B_2)_s
\cr
\pd{0}(A_2)_n-\pd{n}(B_2)_{0}&=-{\e_n}^{ls}\pd{l}(B_1)_s\ .}
\eqn\crk
$$
The solution of these equations linear in velocities is
$$
\eqalign{
(B_1)_0&=\sum_{i=1}^{N}\beta_i {\bf{v}}_i .
 {\bf{w}}({\bf{x}}-{\bf{y}}_i)
\cr
(B_2)_0&=\sum_{i=1}^{N}\alpha_i {\bf{v}}_i .
 {\bf{w}}({\bf{x}}-{\bf{y}}_i)
\cr
{\bf{B_1}}&=\sum_{i=1}^{N}{{\alpha_i} \over |{\bf{x}}-{\bf{y}}_i|}
{\bf{v}}_{i}
\cr
{\bf{B_2}}&=-\sum_{i=1}^{N}{{\beta_i} \over |{\bf{x}}-{\bf{y}}_i|}
{\bf{v}}_{i}\ ,}
\eqn\crl
$$
where ${\bf{w}}$ is the Dirac vector 
potential defined so that
$$
\nabla \wedge {\bf{w}}({\bf{x}})=
 {1 \over |{\bf{x}}|^3}{\bf{x}}\ .
\eqn\uthree
$$

So far our analysis applies to 
all solutions that we have
found in section two. In what 
follows, we shall present 
the  moduli space metric only for those solutions that 
preserve $1/4$ of  spacetime
supersymmetry. The moduli space metric for the rest of
the solutions will be given in the next section.
For the solutions
that preserve $1/4$ of supersymmetry, 
the transverse scalars
$X,Y$ obey the linear relation \lirel. 
Without loss of generality,
we can take $r=0$ to find that
 $\alpha_i=0$. So $X=m_1$ constant and we
choose $X=0$.  It turns out that 
in this case the solution
\crl\ of  \crk\ also solves the {\sl field equations
of the non-linear theory} up to terms linear in velocities.
Moreover, the deformation of the charges \cri\ also obeys
\sfe\ and therefore there are no
 contributions in the effective
action linear in the velocities. 
Substituting our solution
back into the action
$$
S=S_{D3}+S_{Source}+S_{free}\ ,
\eqn\ffone
$$
where
$$
S_{free}= -\sum_{i=1}^N M_i(R) \int d\tau_i+
 4\pi \sigma m \sum_{i=1}^N \beta_i \int
d\tau_i
\eqn\fftwo
$$ 
and collecting the terms
quadratic in the velocities, we find
$$
\eqalign{
S_{Eff}&=\int dt \big({1\over2} 
\sum_{i=1}^{N} M_i(R) |{\bf{v}}_i|^2 -2 \pi
\s\big[\sum_{i=1}^{N} {m \b_i } {\x_i}^2
 +\sum_{i \neq j}^{N} {\b_i \b_j \over r_{ij}}
({\x_i}^2- \x_i \x_j)
\cr
&+\sum_{i \neq j}^{N} {\b_i \b_j \over r_{ij}}
({\bf{v}}_i . {\bf{v}}_j - |{\bf{v}}_i|^2)+2\b_i 
\b_j \x_i ({\bf{v}}_j - {\bf{v}}_i ).{\bf{w}}_{ij}\big]\big)\ ,}
\eqn\crm
$$
where $m=m_2$,
$$
r_{ij}=|{\bf y}_i-{\bf y}_j|\ ,
\eqn\ufour
$$
$\{M_i (R); i=1,\dots, N\}$ is the rest mass 
of the dyons\foot{We have computed
the masses of dyons in the linearized theory.  
Unless  the solution
is appropriately \lq regularized', 
the masses  diverge
in the same way as those of the rest of 
0-brane solitons.}, and ${\bf{w}}_{ij}=
{\bf{w}}({\bf y}_i-{\bf y}_j)$. To 
compute \crm, we remark that
both the Lagrangian density \pola\ 
and the source terms \arg\
contribute. Moreover the second term in 
$S_{free}$ has been added in order that
the effective action of the dyons is 
independent of the asymptotic
position of the D-3-brane as for the 0-branes 
in the previous section. Because $m$ 
appears in the effective Lagrangian above, it 
seems that the moduli metric of the dyons
depends on the position of the D-3-brane. However 
this is not the case since the $m$
dependent terms do not contribute in the field equations and can be tuned
appropriately  to write the Lagrangian in a more symmetric way.

The effective Lagrangian that we have 
obtained above is 
similar, up to adjusting some parameters, to that derived in
[\gima] for well-separated BPS 
monopoles. To continue,  we shall
make use of the reasoning in that paper to find that the 
moduli space of our solutions is a 
toric hyper-K\"ahler manifold.
For this, we first rewrite the Lagrangian of  \crm\  as
$$
L = {1\over2} g_{ij}{\bf{v}}_i {\bf{v}}_j + {1 \over \k}
\chi_i {\bf{W}}_{ij}.{\bf{v}}_j-{1 \over 2 \k^2}h^{ij}\x_i \x_j
\eqn\cro
$$
where $\k$ is a constant, $h^{ij}= {\k^2 }g_{ij}$, 
$$
\eqalign{
g_{jj}&={M_j(R) }+4\pi \s \sum_{i \neq j}^{N} {\b_i \b_j \over r_{ij}}
\cr
g_{ij}&=-4\pi \s{\b_i \b_j \over r_{ij}}\ , \qquad i\not=j\ ,}
\eqn\crp
$$
and
$$
\eqalign{
{\bf{W}}_{jj}&= 4\pi \k  \s \sum_{i \neq j}^{N} \b_i \b_j{\bf{w}}_{ij}
\cr
{\bf{W}}_{ij}&=-4\pi \s \b_i \b_j \k {\bf{w}}_{ij}\ , \qquad i\not=j\ .}
\eqn\crr
$$
We have also tuned appropriately the $m$-dependent terms in \crm. 
The above Lagrangian has equations of 
motion which are identical to those
of
$$
\cL={1\over2}\big[ g_{ij}{\bf{v}}_i {\bf{v}}_j+g^{ij}
({\dot{\tt}}_i +{1 \over \k}{\bf{W}}_{ik}.
{\bf{v}}_k)({\dot{\tt}}_j +{1 \over \k}
{\bf{W}}_{jl}.{\bf{v}}_l)\big]
\eqn\equivb
$$
up to a rescaling of $\tt_i$. To see this, we remark
that 
$$
Q^i=g^{ij} ({\dot{\tt}}_j +{1 \over \k}
{\bf{W}}_{jl}.{\bf{v}}_l)
\eqn\ufive
$$
are conserved subject to field 
equations. To relate the field
equations we simply set $\chi_i= Q^i$.
The coordinates $\theta$ parameterize
the deformations of the charges.

The metric 
$$
ds^2=g_{ij}d{\bf{x}}_i . d{\bf{x}}_j+g^{ij}
(d{{\tt}}_i +{1 \over \k}{\bf{W}}_{ik}.
{d\bf{x}}_k)(d{{\tt}}_j +{1 \over \k}
{\bf{W}}_{jl}.d{\bf{x}}_l)
\eqn\lasttt
$$
 on the  moduli space of dyons given by the effective
Lagrangian \equivb\  is  toric hyper-K\"ahler.
In fact it is a special case of the toric hyper-K\"ahler
metrics of [\ggpt]. In the context of non-abelian
effective theory of D-3-branes, our calculation describes
the  moduli space of well-separated (non-abelian) dyons.
The  effective action we have found is
expected to be that of a one-dimensional sigma model 
with eight supersymmetries.
The geometry we have found is consistent with the
target space geometries of 
 such sigma models (see for example [\cpapad]).
The dyons we have investigated are T-dual to the worldvolume 0-branes
of the D-4-brane [\ppd].
Our results above suggest that this duality 
is extended to their
dynamics. This is because the toric
 hyper-K\"ahler moduli spaces
that we have found here are, under 
sigma model duality, dual to the
hyper-K\"ahler with torsion moduli 
spaces found for the worldvolume
0-brane solitons of D-4-brane in [\jgp]. 
Note that if we neglect the charge moduli parameters
$\x_i$, then we recover the moduli metric
of the 0-brane worldvolume 
solitons on the D-3-brane of [\jgp] (see also [\blla]).

\chapter{Other Dyons}

The computation of the moduli metric 
for the configurations
that preserve $1/8$ of spacetime 
supersymmetry is similar to
that of the dyons that preserve 
$1/4$ of supersymmetry. However
there are some  differences. First the 
solution \crl\ of the linearized
field equations is no longer a 
solution of the non-linear theory
up to terms linear in velocities. In addition, the parameters
do not satisfy the equation \sfe. Consequently, 
one expects that
the effective action should contain linear
terms in the velocities. Moreover in order to obtain
the full effective action quadratic in the 
velocities, quadratic velocity
deformations in the charges should be taken into account.

We have not been able to obtain the 
solution for the deformation
of the one-form gauge potential within 
the non-linear theory. So we
shall use the solution \crl\ to compute 
the moduli metric
for the linearized theory. The effective action is
$$
\eqalign{
\cL_{eff}=&  \int dt \big({1\over2} M_i(R) |{\bf{v}}_i|^2 -2 \pi \s \big[
\sum_{i=1}^{N}-(m_1 \a_i +m_2
\b_i) |{\bf{v}}_i|^2
\cr
+&\sum_{i \neq j}^{N}[{(\a_i \a_j +\b_i \b_j) 
\over r_{ij}}({\x_i}^2-\x_i \x_j) +{(\a_i \a_j
 +\b_i \b_j) \over r_{ij}}({\bf{v}}_i
 \ {\bf{v}}_j - |{\bf{v}}_i|^2)
\cr
+&2(\a_i \a_j +\b_i \b_j)\x_i ({\bf{v}}_j - 
{\bf{v}}_i).{\bf{w}}_{ij}+2(\a_i \b_j - \a_j
 \b_i){\bf{v}}_j . {\bf{w}}_{ij}]
\cr
+&2{(\a_i \b_j - \a_j
 \b_i) \chi_i\over r_{ij}}\big]\big) \ .}
\eqn\dri
$$

The second order part of this Lagrangian is 
 very similar to that in \crm,and it
defines a toric hyper-K\"ahler moduli space
 metric. However, there are additional
first order terms. This effective Lagrangian
 may have additional nonlinear contributions 
added to it when
the full effect of the nonlinear equations
 of motion is taken into account.
The part of our effective action which does 
not involve contributions
from the charge deformation agrees with the one
computed in [\blla] for solutions of 
Dirac-Born-Infeld actions that preserve
$1/8$ of supersymmetry.

\chapter{Conclusions}

We have found the static solutions of the
field equations of a $SL(2,{\bR})$ covariant
D-3-brane action which have the interpretation of dyons
on the D-3-brane worldvolume. We have identified them 
with the dyons generated by a (p,q)-string ending on the
D-3-brane. We have found using $\k$-symmetry that these dyons
preserve   $1/4$  of spacetime supersymmetry. In addition,
we have shown that the moduli space of well-separated dyons 
is a toric
hyper-K\"ahler manifold.  Other 
worldvolume dyons are also found.
We argue that these are associated using 
duality to the M-theory
configuration of two orthogonal 
membranes ending on a M-5-brane.
We have shown that these dyons preserve 
$1/8$ of supersymmetry
and we have computed their effective action.

It would be of interest to have a better description
of the moduli space of all worldvolume 
brane solitons, specially 
those that preserve less than $1/4$ of bulk supersymmetry.
In many cases such description is
 known. For example,
the moduli spaces of worldvolume solitons 
that are associated with calibrations
have been extensively studied in the 
mathematics literature.
The same applies, at least in the linearized limit,  
for the moduli spaces of the
worldvolume solitons that can be 
identified with BPS solitons or instantons.
It would be of interest to investigate 
the moduli spaces of worldvolume
solitons that are associated with both 
calibration-like surfaces and non-abelian
Born-Infeld field configurations.

\vskip 0.5cm

\noindent{\bf Acknowledgments:} We would like to thank A.Westerberg for
many useful discussions. J.G is 
supported by an EPSRC studentship and G.P 
is supported by a University Research Fellowship from the Royal
Society.

\refout
\bye

\refout
\end